\documentclass[reprint,nofootinbib,amsmath,amssymb,prl]{revtex4-2}
\usepackage{dcolumn}
\usepackage{bm}
\usepackage{soulutf8}

\usepackage[english]{babel}
\usepackage[utf8x]{inputenc}
\usepackage[T1]{fontenc}
\usepackage{ulem}

\usepackage[a4paper,top=3cm,bottom=2cm,left=3cm,right=3cm,marginparwidth=1.75cm]{geometry}

\usepackage{graphicx}
\usepackage[center]{caption}
\usepackage{color}
\usepackage[colorlinks=true, allcolors=blue]{hyperref}
\definecolor{nred}{RGB}{224,0,0}
\definecolor{nblue}  {RGB}{28,130,185}
\definecolor{dgreen} {RGB}{78,138,21}
\definecolor{norange}{RGB}{230,120,20}

\begin{document}

\title{The regulator dependence in the functional renormalization group: a quantitative explanation}

\author{Gonzalo De Polsi}
\affiliation{Instituto de F\'isica, Facultad de Ciencias, Universidad de la
Rep\'ublica, Igu\'a 4225, 11400, Montevideo, Uruguay
}%

\author{Nicol\'as Wschebor}
\affiliation{Instituto de F\'isica, Facultad de Ingenier\'ia, Universidad de la
Rep\'ublica, J.H.y Reissig 565, 11000 Montevideo, Uruguay
}%

\date{\today}

\begin{abstract}
  The search of controlled approximations to study strongly coupled systems remains a very general open problem. Wilson's renormalization group has shown to be an ideal framework to implement approximations going beyond perturbation theory. In particular, the most employed approximation scheme in this context, the derivative expansion, was recently shown to converge and yield accurate and very precise results. However, this convergence strongly depends on the shape of the employed regulator. In this article we clarify the reason for this dependence and justify, simultaneously, the most commonly employed procedure to fix this dependence, the principle of minimal sensitivity.
\end{abstract}

\maketitle

\section{Introduction}

When studying strongly coupled physical systems, the search of controlled approximation schemes that enable the computation of quantities of interest is a difficult task. Moreover, as approximations are being made, spurious dependencies on non-physical parameters are usually introduced. In many areas of physics, one of the most popular criteria for fixing this artificial dependence on scheme's parameters is the Principle of Minimal Sensitivity (PMS) \cite{Stevenson1981,Canet2003,Canet:2003qd}. In a nutshell, it consists in taking the non-physical parameters, appearing due to the implementation of a given approximation scheme, at the values at which quantities of interest exhibit the least or locally vanishing dependence.

One of the most efficient methods for treating strongly correlated systems is the Renormalization Group (RG) introduced by Wilson in the 1970s \cite{Wilson1971,Wilson1972}. In particular, its modern version, known as the Non-Perturbative or Functional Renormalization Group (FRG) brought about important progress in the study of critical phenomena. For an extensive review on the subject and its recent results we refer the reader to \cite{Dupuis2021}.

The FRG, being a modern version of Wilson's RG, consists in getting rid of short wavelength fluctuations by integrating them first while the long wavelength fluctuations are kept still. This is done by adding a regulating term $\Delta S_k$ to the action which gives an artificial mass or a finite correlation length to modes with wave number smaller than certain arbitrary scale $k$. Afterwards, this scale $k$ is gradually reduced from the microscopic scale $\Lambda$ at which the theory is defined, until at $k=0$ the original system with all its fluctuations is recovered.
This procedure results in an exact evolution equation for an effective action interpolating between the microscopic action $S$ defined at an ultraviolet scale and the Gibbs free energy or the generating functional of one-particle irreducible correlation functions. However, in order to solve this equation the use of approximation schemes is, in general, unavoidable. One of the most used approximation schemes in this context is known as the Derivative Expansion (DE). This is so for bosonic theories. In the presence of fermionic degrees of freedom typically other kind of approximations are required. To fix ideas we focus here to case of scalar bosonic degrees of freedom. This approximation scheme considers an ansatz for the effective action consisting in an expansion in the number of derivatives acting on the fields. This is very useful because it permits the controlled study of long-distance properties of very general models without having to assume the smallness of any coupling constant. As it was recently shown \cite{Balog2019}, in order for the DE to yield accurate and convergent results that are controlled by a small parameter it is important to pick the profile of the regulating term appropriately. The PMS is a criterion used to fix the regulator dependence and evidence shows that its implementation yields accurate and precise results which improve at successive orders of the DE. One finds that by implementing the PMS error bars are reduced, approximately, by a factor four at each successive order of the DE. Moreover, it was shown recently \cite{Balog2020} that for the critical $\phi^4$ scalar theory, the PMS corresponds to considering a theory satisfying, at the critical point, restrictions coming from conformal invariance as much as possible.

Despite the evidence and the arguments given in the recent past, the PMS criterion is still the weakest point of the DE procedure. This difficulty is no longer a purely academic issue restricted to the community working with FRG methods. In fact, the DE has recently achieved results of great precision and accuracy \cite{Balog2019,DePolsi2020,Peli:2020yiz,DePolsi2021}, in some cases reaching the highest precision in the literature. This promotes it to one of the reference methods in the study of critical phenomena (see, for example \cite{Hasenbusch:2021rse}). These successes make the significative dependence on the regulator not only an internal concern of FRG practitioners but also of those who analyses various problems that go beyond perturbation theory such as critical phenomena (see, for example, \cite{Giuliani2021}). It worth to mention that other optimization criteria have been proposed in order to choose the regulator within FRG \cite{Litim:2000ci,Litim02,Codello_2014,Pawlowski:2015mlf,Braun:2018svj}.

In this article we explain how the regulator should be chosen at large orders of the DE and give evidence that the PMS rapidly converges to the optimum choice. We present first the rudimentary concepts of the FRG that are needed to justify for the appropriate fixing of the regulator at large orders of the DE. We then apply this quantitative estimate to three different families of regulators and compare with the numerical results obtained using the PMS at succeeding orders of the DE, showing that empirical results support our theoretical analysis.

\section{The non-perturbative renormalization group and the asymptotic behavior of the overall regulator scale}\label{secFRGPMS}

As described before, the FRG consists in adding an infrared regulating term to the action $S$ that, when varied, allows to progressively integrate fast fluctuations.
\subsection{The non-perturbative renormalization group in a nutshell}
It is convenient to take it quadratic in the fields \cite{Polchinski1984}: 
\begin{equation}\label{deltaS}
    \Delta S_k[\varphi]=\frac 1 2 \int_{x,y}\varphi_i(x) R_k(x,y) \varphi_i(y),
\end{equation}
where $\int_x=\int d^dx$ and the regulating function $R_k$ (or just the regulator from now on) is picked as to preserve rotation and translation invariance and, therefore, depending only on $|x-y|$ or, otherwise, its Fourier transform is a function of $q^2$. Here and below, Einstein convention is used for repeated internal indices. This is
\begin{equation}\label{regGen}
	R_k(q^2)=\alpha Z_k r_k(q^2),
\end{equation}
where $\alpha$ is the overall scale (which is a real number of order one) of the regulator and $Z_k$ is a field renormalization factor. Additionally, to preserve analyticity in Fourier space, $r_k(q^2)$ should be a smooth function of the momentum $q$ and behave as a mass term of order $k$ \textit{only} for the slow modes, relative to the scale $k$: 
\begin{equation}\label{eqRSM}
    r_k(q^2)\underset{q\to 0}{\sim}k^2 -z q^2+w \frac{q^4}{k^2}+\dots,
\end{equation}
and must vanish sufficiently fast for large momentum ($q^2 \gg k^2$) in order to effectively integrate fast momentum modes. We assume for simplicity, that $w>0$. In equation (\ref{eqRSM}) the coefficients $z$ and $w$ are real numbers of order one that depend on the precise shape of the regulator.

Typical regulators with this behavior used in the literature are the \textit{exponential} regulator:
\begin{equation}\label{regExp}
    r_k(q^2)=E_k=k^2e^{-q^2/k^2};
\end{equation}
the \textit{Wetterich} regulator:
\begin{equation}\label{regWett}
    r_k(q^2)=W_k=\frac{q^2}{e^{q^2/k^2}-1}
\end{equation}
and the \textit{generalized Litim} regulator of order $n$ \footnote{The non-analytic behavior of the generalized Litim regulator can make undefined the DE at large orders (see below).}:
\begin{equation}\label{regLitG}
    r_k(q^2)=\Theta_k^{(n)}=k^2\bigg(1-\frac{q^2}{k^2}\bigg)^n\Theta(1-\frac{q^2}{k^2}).
\end{equation}

All these properties make the regulating term \eqref{deltaS} an infrared regulator that maintains modes with $q\gg k$ unaffected. It has become standard since \cite{Wetterich:1992yh,Ellwanger:1993kk,Morris:1993qb} to add to the action this regulating term and to perform a Legendre transform leading to a scale-dependent generating functional of one-particle irreducible correlation functions $\Gamma_k[\phi]$ or scale dependent \textit{effective action}. 
The evolution of $\Gamma_k$ with the scale $k$ or the RG ``time'' $t=\log(k/\Lambda)$ is given by:
\begin{equation}\label{wettericheq}
\partial_{t}\Gamma_{k}[\phi]=\frac{1}{2}\int_{x,y}\partial_{t}R_{k}(x-y)G_k(x,i;y,i),
\end{equation}
where $G_k(x,i;y,j)$ is the full propagator in an external field, which is defined implicitly by the relation:
\begin{align}
\int_{y}& G_k(x,i;y,n)\left[\frac{\delta^2\Gamma_k}{\delta \phi_n(y)\delta\phi_j(z)}+ R_k(y-z)\delta_{nj}\right]\nonumber\\&=\delta(x-z)\delta_{ij}.
\end{align}
\subsection{Behaviour of regulators in the asymptotic limit of the derivative expansion}
When working with equation \eqref{wettericheq} approximation schemes are generally needed, being the DE one of the most used. It is now well established that the obtained results will depend on the regulator and that this dependence must be fixed by using some criterion. It has also been observed that the main dependence is on the overall scale of the regulator $\alpha$. One such criterion is the PMS and it has been recently used to yield highly accurate and precise results on critical exponents and universal amplitude ratios \cite{Balog2019,DePolsi2020,Peli:2020yiz,DePolsi2021}. Once the dependence on $\alpha$ is fixed by the PMS criterion, the remaining dependence on the different families of regulators turns out to be small \cite{Canet2003,Balog2019,DePolsi2020,DePolsi2021}. Because of this, we focus in the rest of the article on the $\alpha-$dependency but the presented analysis is completely general.

Despite the success of the PMS procedure, its vital relation with the convergence of the DE is not completely elucidated. To solve this difficulty, an estimation can be made on what to expect of the optimum value of the overall regulator scale $\alpha$ at large orders of the DE. To do so, it is possible to set a range of ``applicability'' for the DE based on two arguments. First, the integrand in the flow of the various vertex functions can be obtained from derivatives of equation \eqref{wettericheq}. As a consequence it is proportional to $\partial_t R_k(q^2)$ times the propagator to some power which suppress large momenta. In general it takes the following form:
\begin{equation}
\int_q \partial_t R_k(q^2) \Big(G(q^2)\Big)^n (q^2)^m,
\end{equation}
(in some cases some propagators are differentiated one or more times).
The contribution to these integrals of internal momenta $q$ larger than, roughly speaking, $q^2\gtrsim k_{cut}^2$ are negligible, where $k_{cut}^2$ is proportional to $k^2$, say
\begin{equation}
k_{cut}^2=k^2/\beta,
\end{equation}
with $\beta$ a real number of order one whose precise determination is not trivial. We discuss how to estimate it below but, for the moment, let us keep the value of $\beta$ general because some of our results do not depend on its precise value.  It is important to stress that $R_k(q^2)$ does not only introduce a gap $\propto \alpha k^2$ but also modifies the normalization of the kinetic term $\propto q^2$ in the inverse propagator. Taking this into account, 
\begin{equation}
\Gamma^{(2)}(q^2)+R_k(q^2)\cong Z_k\Big(\alpha k^2+ q^2 (1-\alpha z)+\mathcal{O}(q^4)\Big), 
\end{equation}
at small $q$. Accordingly, the mass gap can be readily obtained: 
\begin{equation}\label{eqMassGap}
    m_{gap}^2=\frac{\alpha k^2}{1-\alpha z}.
\end{equation}
In the previous expression it has been assumed that $|\Gamma_k^{(2)}(q^2=0)|\ll R_k(q^2=0)$ as it is usually the case in critical phenomena. We also assumed that $\eta$ is relatively small in order to neglect the dependence on $\phi$
in $\frac{d}{d q^2}\Gamma_k^{(2)}|_{q^2=0}$. The massive behavior of the regulated theory implies that it exhibits a radius of convergence in momentum regardless of the theory being in the broken or the symmetric phase. The radius of convergence is given by the closest singularity to the origin in the complex $p^2$ plane and for the symmetric phase this is located at $p^2=-9m^2$ and at $p^2=-4m^2$ in the broken phase 
(see, for example, \cite{Pelissetto:2000ek}). As discussed in \cite{Balog2019}, these behaviors follow from the fact that 
the regulated theory behaves as a massive theory for small momenta. In the massive theory, as is well-known, the closest singularity to the origin in the complex $p^2$ plane of the momenta dependence of connected correlation functions are located at $p^2=-m^2$. This singularity corresponds to the single particle pole in the Minkowskian version of the theory. In the 1-PI case, this singularity is no longer present and the the singularity nearest to the origin in the complex $p^2$ plane is at $-9 m^2$ ($-4m^2$) because, again, the Minkowskian extension of the theory has a three- (two-) particle branch-cut in the symmetric (broken)
phase respectively.

Consequently, when the vertex functions appearing in the right hand side integrals of flow equations are expanded both in internal and external momenta, the radius of convergence is $4 m^2_{gap}$ (using the least favorable case). Now, external momenta can chosen arbitrarily small and internal momenta are bounded by $q\lesssim k_{cut}$. One concludes that in presence of the regulator an expansion in $p^2/m_{gap}^2$ (where $p$ can be external or internal) will present the following expansion parameter:
\begin{equation}\label{eqLambda1}
    \lambda_1=\frac{k_{cut}^2}{4m_{gap}^2}=\frac{1-\alpha z}{4\alpha \beta},
\end{equation}
and one expects the DE to converge when $\lambda_1\lesssim 1$\footnote{At a more technical level, the previous reasoning also explains why the DE is not so well behaved when considered for the Polchinski equation instead of the Wetterich equation because in the former case connected 1-particle reducible correlation functions contribute, and for the connected correlation function the closest singularity to the origin in the $p^2$ plane is then situated at $p^2=-m^2$.}. The previous reasoning express the ideas developed in Ref.~\cite{Balog2019} but incorporating the dependence on $\alpha$ of the various quantities. It is interesting to note that the original idea of maximizing the gap proposed by Litim\cite{Litim:2000ci} is similar to maximizing $\lambda_1$ and gives similar results. However, improving $\lambda_1$ alone can not be the end of the story because it is known that even if the Litim's regulator gives the better results at order LPA, the DE becomes ill-defined at higher orders with this specific regulator.

A second condition needs to be satisfied as well. Indeed, since the previous reasoning suppose that the theory in presence of the regulator must behave approximately as a massive theory in the range $q^2\lesssim k_{cut}^2$, the contribution in this range from higher order terms in \eqref{eqRSM} must be small in comparison with quadratic terms. This means that we can identify another parameter $\lambda_2$ in the expansion given by the ratio of the quartic and quadratic terms in momentum in the inverse propagator \footnote{Similarly, one could consider other parameters coming from the comparison of higher powers of momenta with the quadratic term.}:
\begin{equation}\label{eqLambda2}
    \lambda_2=\frac{\alpha w k_{cut}^4/k^2}{(1-\alpha z)k_{cut}^2}=\frac{\alpha w}{\beta(1-\alpha z)}.
\end{equation}

Because of the nature of these two parameters, $\lambda_1$ and $\lambda_2$, a range of validity of the DE can be established at large orders of the DE from the conditions:
\begin{subequations}\label{conditions}
	\begin{align}
		\label{condlambda1}\lambda_1(\alpha)&=\frac{1-\alpha z}{4\alpha \beta}\lesssim 1,\\
		\label{condlambda2}\lambda_2(\alpha)&=\frac{\alpha w}{\beta(1-\alpha z)}\lesssim 1.
	\end{align}	
\end{subequations}
It is straightforward to see that these conditions read:
\begin{equation}\label{eqAlphaRange}
    \alpha_{min}\equiv\frac{1}{z+4\beta}\lesssim\alpha\lesssim\frac{\beta}{w+z\beta}\equiv\alpha_{max}.
\end{equation}

\begin{figure}[!ht]
	\includegraphics[width=\linewidth]{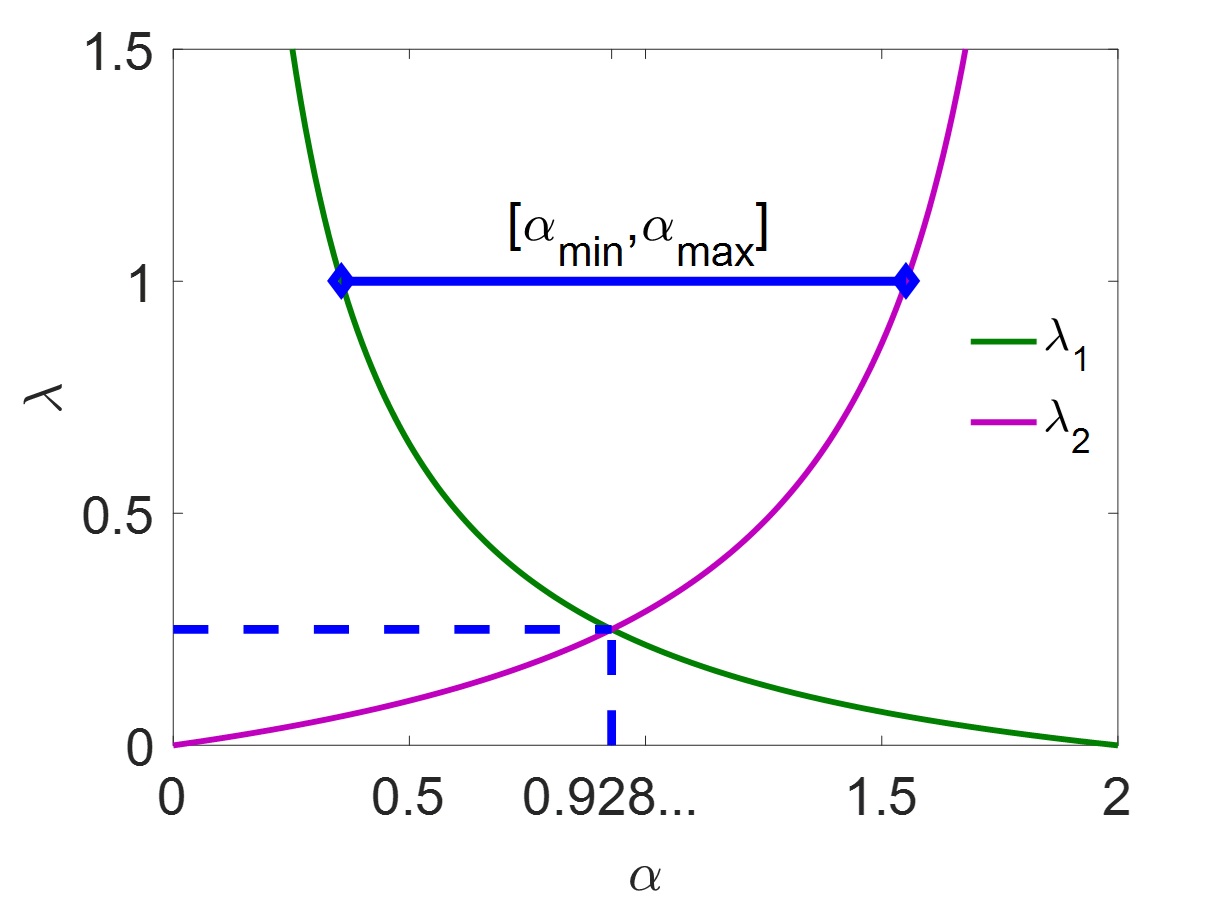}
	\caption{\label{Fig_Lambdas} Dependence of $\lambda_1$ and $\lambda_2$ as a function of $\alpha$ for Wetterich regulator $W_k$ with $\beta=2\sqrt{w}$. The crossing of the curves for $\lambda_1$ and $\lambda_2$ denotes $\lambda_{opt}$ located at $\alpha_{opt}$ and a
		lso the range given by $[\alpha_{min},\alpha_{max}]$ is highlighted (blue).}
\end{figure}

Moreover, once again based on the nature of these parameters, one could infer an \textit{optimal} value $\alpha_{opt}$ by considering the case where both expansion parameters are as small as possible (see Fig.~\ref{Fig_Lambdas}). Equivalently, we require both values to be equal 
\begin{equation}
\lambda_1(\alpha_{opt})=\lambda_2(\alpha_{opt})=\lambda_{opt}. 
\end{equation}
A straightforward calculation yields:
\begin{equation}\label{eqAlphaOpt}
    \alpha_{opt}=\frac{1}{z+2\sqrt{w}},
\end{equation}
corresponding to a parameter:
\begin{equation}\label{eqLambdaOpt}
    \lambda_{opt}=\frac{\sqrt{w}}{2\beta}.
\end{equation}
It is interesting to observe that the value of $\alpha_{opt}$ does not depend
on $\beta$. This makes its prediction very robust. At odds with that, the value of $\lambda_{opt}$ does depend on $\beta$. This parameter can be estimated by looking at the point where the expansion (\ref{eqRSM}) leads to a
\begin{equation}
\partial_t R_k(q^2)= \alpha Z_k \big( k^2(2-\eta_k)+\eta_k z q^2-2 w \frac{q^4}{k^2}+\dots\big)
\end{equation}
that extrapolates to zero. Here, as usual, $\eta_k=-\partial_t \log Z_k$. When neglecting $\eta_k$ this value can be estimated easily to be $q^2=k^2/\sqrt{w}$. Accordingly, one sees that, as long as the expression (\ref{eqRSM}) extrapolates smoothly until $q^2\sim k_{cut}^2$ one expects $k_{cut}^2\sim k^2/\sqrt{w}$ or, equivalently, $\beta\sim \sqrt{w}$. In fact, the precise coefficient in front of $\sqrt{w}$ is somewhat arbitrary as it depends on which part of the tail of the integrals one disregards. Moreover, the actual integrals include the propagator to some powers which furthermore suppress higher momenta. In order to make things concrete, we will take from now on $\beta=2\sqrt{w}$. We verified by a specific study of some integrals appearing in RG equations that, indeed a $\beta$ of that order gives rise to a reasonable cut-off point for the integrals. This value of $\beta$ corresponds to $\lambda_{opt}$ which, moreover, turns out to be roughly the observed expansion parameter in concrete DE studies. A byproduct of this analysis is that the value of the expansion parameter turns out to be insensitive to the families of regulator once the value of $\alpha$ is optimized. 

According to the previous discussion, the values of $z$ and $w$ as well as the expected values of $\alpha_{opt}$, $\alpha_{min}$ and $\alpha_{max}$ are presented in Table \ref{TabAlphRegs} for the four families of regulators given in equations \eqref{regExp}, \eqref{regWett} and \eqref{regLitG} with $n=3$ and $n=4$. For completeness, Fig.~\ref{Fig_Lambdas} illustrates the arguments for the case of the Wetterich regulator \eqref{regWett}.

\begin{table}[!h]
		\begin{tabular}{c|cccccc}
			$\mathcal{R}_k$ & $z$ & $w$ & $A_\mathcal{R}$ & $\alpha_{opt}$ & $\alpha_{min}$ & $\alpha_{max}$ \\\hline
			$E_k$ & $1$ & $1/2$ & $0.487$ & $0.414$ & $0.150$ & $0.739$ \\
			$W_k$ & $1/2$ & $1/12$ & $1.072$ &  $0.928$ & $0.356$ & $1.552$ \\
			$\Theta^{(3)}_k$ & $3$ & $3$ & $-$ & $0.155$ & $0.059$ & $0.259$ \\
			$\Theta^{(4)}_k$ & $4$ & $6$ & $0.142$ & $0.112$ & $0.042$ & $0.191$ \\
		\end{tabular}\caption{\label{TabAlphRegs} Values of $z$, $w$, $A_\mathcal{R}$, $\alpha_{opt}$ $\alpha_{min}$ and $\alpha_{max}$ for the $E_k$, $W_k$, $\Theta_k^{(3)}$ and $\Theta_k^{(4)}$ regulators computed using equations \eqref{eqAlphaRange}, \eqref{eqAlphaOpt} and \eqref{eqLambdaOpt}. The values presented for $A_{\mathcal{R}}$ are fitted estimations using as reference the values obtained for the critical exponent $\nu$ of the $N=1$ case and it is only estimated for the $E_k$, $W_k$ and $\Theta_k^{(4)}$ regulators computed using equations \eqref{eqAlphaRange}. Notice that the value of $\lambda_{opt}$ is always $1/4$ due to our estimate of $\beta$.}
\end{table}


\begin{figure*}[!t]
	\centering
	\includegraphics[width=\textwidth]{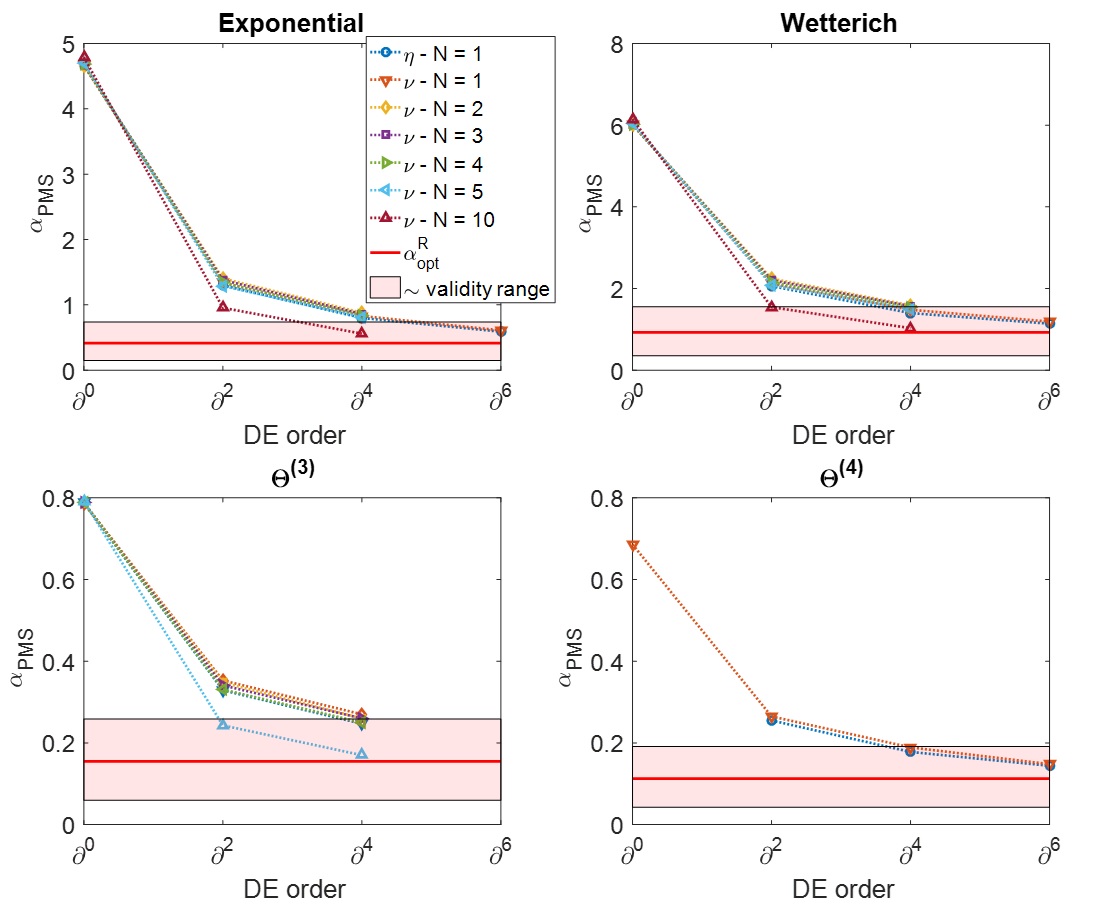}
	\caption{\label{Fig_PMSEtaNu}Behavior of PMS values of $\alpha$ for critical exponents $\eta$ and $\nu$ and different models as a function of the order of the DE. Optimum value $\alpha_{opt}$ (see Eq.~\eqref{eqAlphaOpt}) is shown with a red solid line and the approximate range of applicability given by $\alpha_{min}$ and $\alpha_{max}$ obtained with equation \eqref{eqAlphaRange}. The data correspond to the exponential (top-left), Wetterich (top-right), generalized Litim with $n=3$ (bottom-left) and generalized Litim with $n=4$ (bottom-right).}
\end{figure*}
\begin{figure*}[!t]
	\centering
	\includegraphics[width=\textwidth]{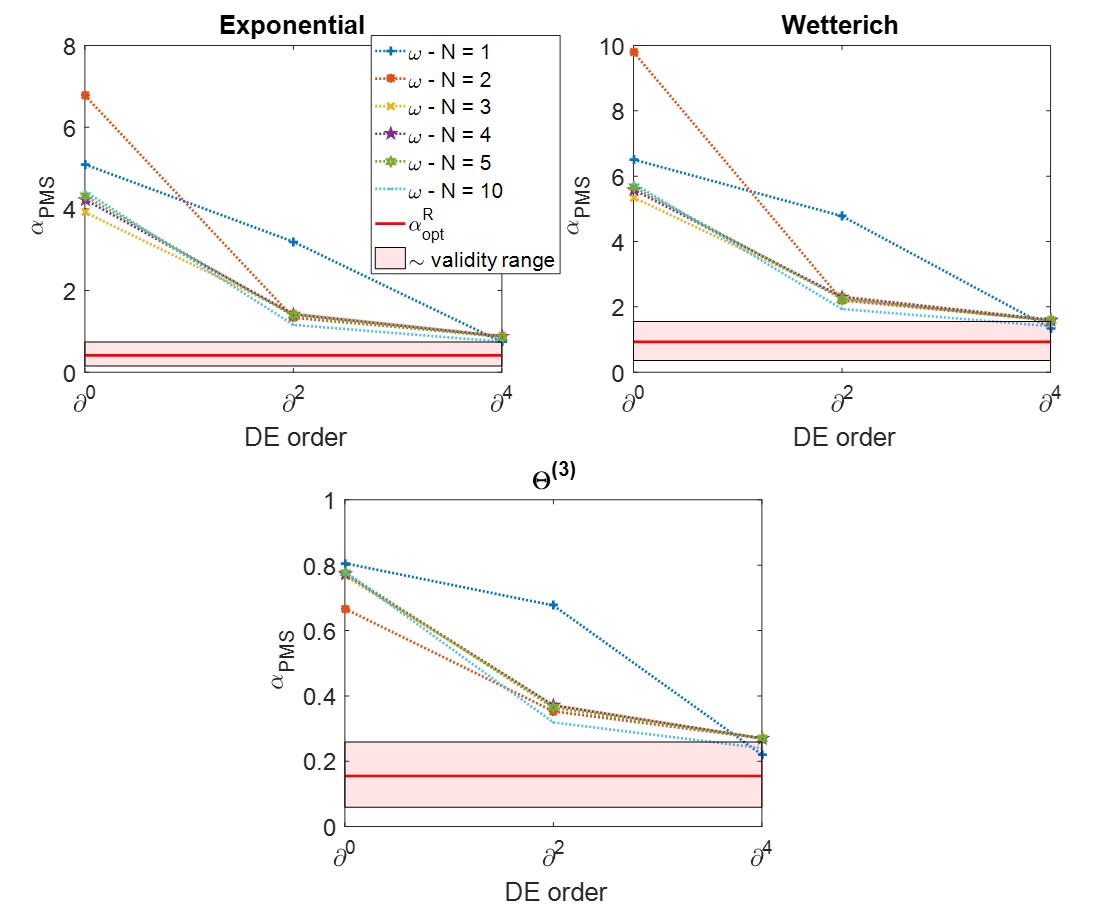}
	\caption{\label{Fig_PMSOmega}Behavior of PMS values of $\alpha$ for the critical exponent $\omega$ for different models as a function of the order of the DE. Optimum value $\alpha_{opt}$ (see Eq.~\eqref{eqAlphaOpt}) is shown with a red solid line and the approximate range of applicability given by $\alpha_{min}$ and $\alpha_{max}$ obtained with equation \eqref{eqAlphaRange}. The data correspond to the exponential (top-left), Wetterich (top-right) and generalized Litim with $n=3$ (bottom).}
\end{figure*}

\subsection{Discarding regulators}
The previous analysis permits to discard many regulators which will not behave properly at arbitrary large orders of the DE. In order to ensure the existence of a small parameter associated to the DE, one needs the Taylor expansion of the regulator to describe it properly until $q^2\approx 4 k_{cut}^2$. This is not so, for example, for a sharp regulator (where the Taylor expansion does not exist) or for a pure mass regulator (constant) because in that case $\partial_t R_k(q)$ does not go to zero and the notion of $k_{cut}$ becomes ill-defined. It may seem that the present analysis is not applicable to these type of regulator either because they are non-analytic or because its Taylor expansion does not allow such an analysis until $k_{cut}$ (consider for example the mass regulator). However, as we discuss below, these regulators can be related to analytic regulators decreasing at large $q$ in certain limit of its parameters.

The sharp regulator and the mass regulator are defined by:
\begin{equation}\label{sharpReg}
	r_k(q^2)=S^{(\infty)}_k(q^2)=\bigg\lbrace\begin{matrix}
	\infty && q^2<k^2\\
	0 && q^2>k^2
	\end{matrix}
\end{equation}
and
\begin{equation}\label{massReg}
		r_k(q^2)=M_k(q^2)=k^2,
	\end{equation}
respectively, according to equation \eqref{regGen}.  We start by considering the \textit{Fermi} regulator $F_k^{(T)}$ dependent on a parameter $T$ defined as:
\begin{equation}\label{FermiReg}
		r_k(q^2)=F_k^{(T)}(q^2)=\frac{k^2}{T^2} \frac{1}{1+e^{\frac{1}{T}(q^2/k^2-1)}},
\end{equation}
It is thus clear that $$\displaystyle{\lim_{T \to 0}}\mathcal{F}_k^{(T)}=S_k^{(\infty)}.$$ At this point we notice that replacing $T$ by $\alpha$ in \eqref{FermiReg}, we would have to consider the limit of $\alpha\to0$. However, it is important to notice that we want the sharp regulator to be infinite for $q^2<k^2$ even when taking $\mathcal{R}_k$ given by \eqref{regGen}. This is the reason to include a factor $T^{-2}$ in \eqref{FermiReg} in order to retain the sharp regulator in the limit ($T=\alpha\to 0$) instead of just $T^{-1}$ which would not have the appropriate limiting behavior.

We also notice that at very small values of $T$ (i.e. in the limit $T\to 0$), it is $k_{cut}=k$ for the Fermi regulator. Therefore, one finds that $$\lambda_1=\frac{1-\alpha z}{4 \alpha}.$$ Now, it is found that in the limit $\alpha\to 0$ it is $z=0$. Consequently, this implies that this regulator violate the condition \eqref{condlambda1} and, therefore, it is not expected to behave properly at arbitrary large orders of the DE.

For the mass regulator, we notice that every smooth regulator will behave, in the limit $\alpha\to0$ as the mass regulator \eqref{massReg}. Indeed, consider the general expression
\begin{equation}\label{genReg2}
	\mathcal{R}_k(q^2)=\alpha k^2 Z_k f(q^2/k^2),
\end{equation}
where $k^2 f(q^2/k^2)=r_k(q^2)$. Now, if 
we redefine $k'^2=\alpha k^{2-\eta}$, 
equation \eqref{genReg2} takes the form:
\begin{equation}
\mathcal{R}_{k'}(q^2)=k'^2 f\left(\alpha^{2/(2-\eta)} \frac{q^2}{k'^{4/(2-\eta)}}\right).
\end{equation}
If now we take the limit $\alpha\to0$ at fixed $k'$ we see that we see that the regulators behave exactly as a constant term for (this is, $q$-independent) as is the case for the mass regulator. As a consequence, we see that the behavior of the mass regulator corresponds to the situation of violating \eqref{condlambda1} by taking the limit $\alpha\to0$.

This shows how the present analysis permits to discard those regulators previously known to behave poorly (see, for example, \cite{Litim02}).

\section{Behavior of PMS values at succeeding orders of the DE}\label{secRes}

As previously described, at large orders of the DE not every value of $\alpha$ will yield a reasonable behavior. This is due to a competition between two different elements. This argument in addition to apply to large orders of the DE, also gives indications for why there is a well behaved region where a diminished sensitivity to the regulator is expected. Indeed, the given argument shows that considering small $\alpha$ values suits one of the conditions but spoils the other one. Similarly, considering large values of $\alpha$ do the opposite. In this sense, when a compromise is taken considering both elements, the results are expected to be rather insensitive to the regulator. As a consequence, the PMS becomes the way of identifying the regime where both elements are best satisfied at a given order.

These ideas can be tested quantitatively by comparing to actual calculations. It is expected that the values of $\alpha$ at which a weaker dependence of the results is exhibited will converge to the values given by equation \eqref{eqAlphaOpt}. To estimate the behavior of $\alpha_{opt}(s)$, this is the optimum value of $\alpha$ at any given order $\mathcal{O}(\partial^s)$ of the DE, one can exploit the fact that, near the optimum value, one expects that any small momentum quantity (including $\alpha_{opt}$ for the various critical exponents) should converge to its asymptotic value by reducing its error by, approximately, $1/4$ at each order. As a consequence, one can estimate the optimum value of $\alpha$ at finite orders of the DE as:
\begin{equation}\label{eqConvPMS}
    \alpha_{opt}(s)\approx A_{\mathcal{R}}+B_\mathcal{R} \bigg(\frac{1}{4}\bigg)^{(s+2)/2},
\end{equation}
where $s$ is the order of the DE, $A_{\mathcal{R}}$  and $B_{\mathcal{R}}$ are parameter that depend on the specific family of regulator being considered. In this work, the three reasonably well-behaved families of regulators \eqref{regExp}, \eqref{regWett} and \eqref{regLitG} are examined and we compare the $\alpha_{PMS}$ obtained at various orders to the approach to the asymptotic value of the $\alpha_{opt}(s)$ given by (\ref{eqConvPMS}). Before fitting the data, please note that although $A_\mathcal{R}$ is not expected to depend on the model being considered, the values of $B_{\mathcal{R}}$  would probably do. The fitting\footnote{We also tried other fittings such as considering the small parameter to vary instead of being fixed at $1/4$ while fixing $A_\mathcal{R}=\alpha_{opt}$. All the fittings display essentially the same behavior and yield the same conclusions.} with equation \eqref{eqConvPMS} is done only for the values corresponding to the critical exponent $\nu$ for the Ising model with the $E_k$, $W_k$ and $\Theta_k^{(4)}$ regulators where results are available up to order $\mathcal{O}(\partial^6)$ of the DE, meaning that there are 4 points available for fitting.\footnote{The first order $\mathcal{O}(\partial^0)$ of the DE, commonly known as the \textit{Local Potential Approximation}, imposes $\eta=0$ and therefore, no data of PMS behavior is available at that order for the field anomalous dimension $\eta$. Additionally, the behavior of $\Theta_k^{(3)}$ at order $\mathcal{O}(\partial^6)$ is not well defined due to non-analyticities.} The resulting values for $A_\mathcal{R}$ are summarized in Table \ref{TabAlphRegs} and compared to the expected optimum value $\alpha_{opt}$ and range.

It must be emphasized that this fitting is not expected to be very precise as the available data are limited. Moreover, this is the expected behavior at large orders of the DE and the present analysis considers in the same footing the orders $\mathcal{O}(\partial^0)$ and $\mathcal{O}(\partial^6)$. However, it gives a clear picture of agreement with the proposed argument. Additionally, and for visualization, the behavior of PMS values for the critical exponents $\eta$ and $\nu$ is presented in Fig.~\ref{Fig_PMSEtaNu} and for the critical exponent $\omega$ in Fig.~\ref{Fig_PMSOmega} for various $O(N)$ models (including the Ising model $N=1$) as a function of the order of the DE. Despite the fact that a fit for the $O(N)$ models (or for the critical exponent $\eta$) is not reliable due to the small amount of available data, the  behavior shown in Figs.~\ref{Fig_PMSEtaNu}-\ref{Fig_PMSOmega} by these quantities is clearly in line with the previous discussion. Moreover, it can be observed that $A_{\mathcal{R}}$ is somewhat model-independent as our estimate suggests. We must also mention that, as was discussed in \cite{DePolsi2020}, the behavior of the critical exponent $\omega$ is not captured as good as the other critical exponents at small values of $N$, in particular with $N=1$. This is why in Fig.~\ref{Fig_PMSOmega} the ``curve'' of $\omega$ as a function of the order seems not to behave in the same way as the other quantities. Nonetheless we observe also for this exponent an approach to the expected region of validity of the DE when the order of the expansion grows.

It is worth making a final comment of a slightly more technical nature. The analysis just presented also explains qualitatively why at low orders of the DE larger $\alpha_{PMS}$ values are observed and that these values tend to converge progressively to our large-order estimate. Indeed, the $\lambda_2$ parameter compares terms with four derivatives with terms with two derivatives. That is, it is sensitive to contributions that play a minor role in the first orders of the DE. Disregarding the condition on $\lambda_2$ by merely analyzing $\alpha$ values favorable for minimizing $\lambda_1$ tends to favor larger $\alpha$ values, as emerges from empirical calculations.

\section{Conclusions}\label{secConcl}

In summary, a concrete and concise argument has been given regarding the presence and nature of the PMS in the context of the FRG. On top of this, specific values for the overall regulator scale at asymptotically large orders of the DE were given for various families of regulators and the data at finite orders exhibit a convergent behavior toward the expected values. The present analysis focus on the DE of scalar models but it may be generalized to other cases. In particular, it applies almost identically to the BMW approximation scheme \cite{Blaizot:2005xy,Benitez:2011xx} to the same kind of models. Fermionic models requires a separate discussion. These results and the argument previously given put the principle of minimal sensitivity onto firm ground and highlight the importance of its implementation. Moreover, this work gives a preliminary analysis of what to expect from a regulator and where to scan for the overall scale regulator parameter that could be put forward before diving into an extensive and overwhelming calculation.

\acknowledgements

The authors thank B. Delamotte, L. Canet, G. Tarjus and P. Jakubczyk for fruitful comments concerning a previous version of the present manuscript and to A. Codello for useful discussions. We thank PEDECIBA (Programa de desarrollo de las Ciencias B\'asicas, Uruguay). The authors thank the funding of ANII (Uruguay) from grant of code FCE\_1\_2021\_1\_166479.

\bibliographystyle{unsrt}
\bibliography{articuloPMSConv}

\begin{thebibliography}{10}

\bibitem{Stevenson1981}
P.~M. Stevenson.
\newblock Optimized perturbation theory.
\newblock {\em Phys. Rev. D}, 23:2916--2944, Jun 1981.

\bibitem{Canet2003}
L\'eonie Canet, Bertrand Delamotte, Dominique Mouhanna, and Julien Vidal.
\newblock Optimization of the derivative expansion in the nonperturbative
  renormalization group.
\newblock {\em Phys. Rev. D}, 67:065004, Mar 2003.

\bibitem{Canet:2003qd}
Leonie Canet, Bertrand Delamotte, Dominique Mouhanna, and Julien Vidal.
\newblock {Nonperturbative renormalization group approach to the Ising model: A
  Derivative expansion at order partial**4}.
\newblock {\em Phys. Rev. B}, 68:064421, 2003.

\bibitem{Wilson1971}
Kenneth~G. Wilson.
\newblock Renormalization group and critical phenomena. i. renormalization
  group and the kadanoff scaling picture.
\newblock {\em Phys. Rev. B}, 4:3174--3183, Nov 1971.

\bibitem{Wilson1972}
Kenneth~G. Wilson and Michael~E. Fisher.
\newblock Critical exponents in 3.99 dimensions.
\newblock {\em Phys. Rev. Lett.}, 28:240--243, Jan 1972.

\bibitem{Dupuis2021}
N.~Dupuis, L.~Canet, A.~Eichhorn, W.~Metzner, J.M. Pawlowski, M.~Tissier, and
  N.~Wschebor.
\newblock The nonperturbative functional renormalization group and its
  applications.
\newblock {\em Physics Reports}, 910:1--114, 2021.
\newblock The nonperturbative functional renormalization group and its
  applications.

\bibitem{Balog2019}
Ivan Balog, Hugues Chat\'e, Bertrand Delamotte, Maroje
  Marohni\ifmmode~\acute{c}\else \'{c}\fi{}, and Nicol\'as Wschebor.
\newblock Convergence of nonperturbative approximations to the renormalization
  group.
\newblock {\em Phys. Rev. Lett.}, 123:240604, Dec 2019.

\bibitem{Balog2020}
Ivan Balog, Gonzalo De~Polsi, Matthieu Tissier, and Nicol\'as Wschebor.
\newblock Conformal invariance in the nonperturbative renormalization group: A
  rationale for choosing the regulator.
\newblock {\em Phys. Rev. E}, 101:062146, Jun 2020.

\bibitem{DePolsi2020}
Gonzalo De~Polsi, Ivan Balog, Matthieu Tissier, and Nicol\'as Wschebor.
\newblock Precision calculation of critical exponents in the o(n) universality
  classes with the nonperturbative renormalization group.
\newblock {\em Phys. Rev. E}, 101:042113, Apr 2020.

\bibitem{Peli:2020yiz}
Zolt\'an P\'eli.
\newblock {Derivative expansion for computing critical exponents of $O(N)$
  symmetric models at next-to-next-to-leading order}.
\newblock {\em Phys. Rev. E}, 103(3):032135, 2021.

\bibitem{DePolsi2021}
Gonzalo De~Polsi, Guzm\'an Hern\'andez-Chifflet, and Nicol\'as Wschebor.
\newblock Precision calculation of universal amplitude ratios in o($n$)
  universality classes: Derivative expansion results at order
  $\mathcal{O}({\ensuremath{\partial}}^{4})$.
\newblock {\em Phys. Rev. E}, 104:064101, Dec 2021.

\bibitem{Hasenbusch:2021rse}
Martin Hasenbusch.
\newblock {Three-dimensional $O(N)$-invariant $\phi^4$ models at criticality
  for $N\ge 4$}.
\newblock 12 2021.

\bibitem{Giuliani2021}
Alessandro Giuliani, Vieri Mastropietro, and Slava Rychkov.
\newblock {Gentle introduction to rigorous Renormalization Group: a worked
  fermionic example}.
\newblock {\em JHEP}, 01:026, 2021.

\bibitem{Litim:2000ci}
Daniel~F. Litim.
\newblock {Optimization of the exact renormalization group}.
\newblock {\em Phys. Lett. B}, 486:92--99, 2000.

\bibitem{Litim02}
Daniel~F. Litim.
\newblock Critical exponents from optimised renormalisation group flows.
\newblock {\em Nucl. Phys. B}, 631:128--158, 2002.

\bibitem{Codello_2014}
Alessandro Codello, Maximilian Demmel, and Omar Zanusso.
\newblock Scheme dependence and universality in the functional renormalization
  group.
\newblock {\em Physical Review D}, 90(2), jul 2014.

\bibitem{Pawlowski:2015mlf}
Jan~M. Pawlowski, Michael~M. Scherer, Richard Schmidt, and Sebastian~J. Wetzel.
\newblock {Physics and the choice of regulators in functional renormalisation
  group flows}.
\newblock {\em Annals Phys.}, 384:165--197, 2017.

\bibitem{Braun:2018svj}
Jens Braun, Marc Leonhardt, and Jan~M. Pawlowski.
\newblock {Renormalization group consistency and low-energy effective
  theories}.
\newblock {\em SciPost Phys.}, 6(5):056, 2019.

\bibitem{Polchinski1984}
Joseph Polchinski.
\newblock Renormalization and effective lagrangians.
\newblock {\em Nuclear Physics B}, 231(2):269--295, 1984.

\bibitem{Wetterich:1992yh}
Christof Wetterich.
\newblock {Exact evolution equation for the effective potential}.
\newblock {\em Phys. Lett.}, B301:90--94, 1993.

\bibitem{Ellwanger:1993kk}
U.~Ellwanger.
\newblock {Collective fields and flow equations}.
\newblock {\em Z. Phys.}, C58:619--627, 1993.

\bibitem{Morris:1993qb}
Tim~R. Morris.
\newblock {The Exact renormalization group and approximate solutions}.
\newblock {\em Int. J. Mod. Phys.}, A9:2411--2450, 1994.

\bibitem{Pelissetto:2000ek}
Andrea Pelissetto and Ettore Vicari.
\newblock {Critical phenomena and renormalization group theory}.
\newblock {\em Phys. Rept.}, 368:549--727, 2002.

\bibitem{Blaizot:2005xy}
J.~P. Blaizot, Ramon Mendez~Galain, and Nicolas Wschebor.
\newblock {A New method to solve the non perturbative renormalization group
  equations}.
\newblock {\em Phys. Lett.}, B632:571--578, 2006.

\bibitem{Benitez:2011xx}
F.~Benitez, J.~P. Blaizot, H.~Chate, B.~Delamotte, R.~Mendez-Galain, and
  N.~Wschebor.
\newblock {Non-perturbative renormalization group preserving full-momentum
  dependence: implementation and quantitative evaluation}.
\newblock {\em Phys. Rev.}, E85:026707, 2012.

\end{thebibliography}

\end{document}